\documentclass[twoside,12pt]{article}
\usepackage{epsfig}

\def\Journal#1#2#3#4{{#1} {#2} (#4) #3 }

\begin{document}
\title{{\vskip -1cm\normalsize\rm\hfill DESY 05-234}
\vskip 1cm
Strongly Interacting Astrophysical Neutrinos\footnote{Talk presented at the
  Erice School/Workshop on Neutrinos, September 16-24, 2005}}
\author{Markus~Ahlers, DESY Hamburg, Germany}
\date{}
\maketitle

\begin{abstract} 
The origin and chemical composition of ultra high energy cosmic rays is still an open question in astroparticle physics. The observed large-scale isotropy and also direct composition measurements can be interpreted as an extragalactic proton dominance above the {\it ankle} at about $10^{10}$ GeV. Photopion production of extragalactic protons in the cosmic microwave background predicts a cutoff at about $5\times10^{10}$ GeV in conflict with excesses reported by some experiments. In this report we will outline a recent statistical analysis~\cite{Ahlers:2005zy} of cosmic ray data using strongly interacting neutrinos as primaries for these excesses. 
\end{abstract}

\section{Introduction}
There are reasons to believe that cosmic rays (CRs) around the ankle at $10^{10}$ GeV are dominated by extragalactic protons~\cite{Berezinsky:2002nc,Ahlers:2005sn}. Scattering processes in the cosmic microwave background (CMB) limit the propagation of ultra high energy (UHE) charged particles in our Universe. A continuation of a power-like CR spectrum above the Greisen-Zatsepin-Kuzmin (GZK) cutoff~\cite{GZK} at about $5\times10^{10}$ GeV is only consistent with the proton dominance if the sources lie within the proton attenuation length of about 50 Mpc. Very few astrophysical accelerators can generate CRs with energies above the GZK cutoff~(see e.g.~\cite{Torres:2004hk} for a review) and so far none of the candidate sources have been confirmed in our local environment. It has been speculated that decaying superheavy particles, possibly some new form of dark matter or remnants of topological defects, could be a source of UHE CRs, but also these proposals are not fully consistent with the CR spectrum at lower energies~\cite{Semikoz:2003wv}. 

\begin{figure}[t]
\begin{minipage}[t]{\linewidth}
\center
\includegraphics[width=0.5\linewidth]{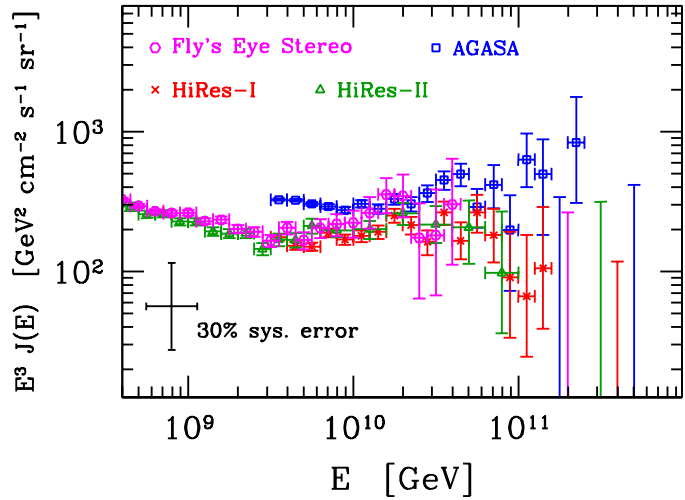} 
\caption{The CR spectrum observed by various experiments.}
\label{CR}
\end{minipage}
\end{figure}

The observation of GZK excesses has led to speculations about a different origin of UHE CRs. Berezinsky and Zatsepin~\cite{Beresinsky:1969qj} proposed that {\it cosmogenic} neutrinos produced in the decay of the GZK photopions could explain these events assuming a strong neutrino nucleon interaction. 
We have followed this idea in Ref.~\cite{Ahlers:2005zy} and investigated the statistical goodness of scenarios with strongly interacting neutrinos from optically thin sources using CR data from AGASA~\cite{Takeda:2002at} and HiRes~\cite{HIRES} (see Fig.~\ref{CR}) and limits from horizontal events at AGASA~\cite{Yoshida:2001pw} and contained events at RICE~\cite{Kravchenko:2003tc}.

\begin{figure}[t]
\begin{minipage}[t]{\linewidth}
\center
\includegraphics[width=0.8\linewidth]{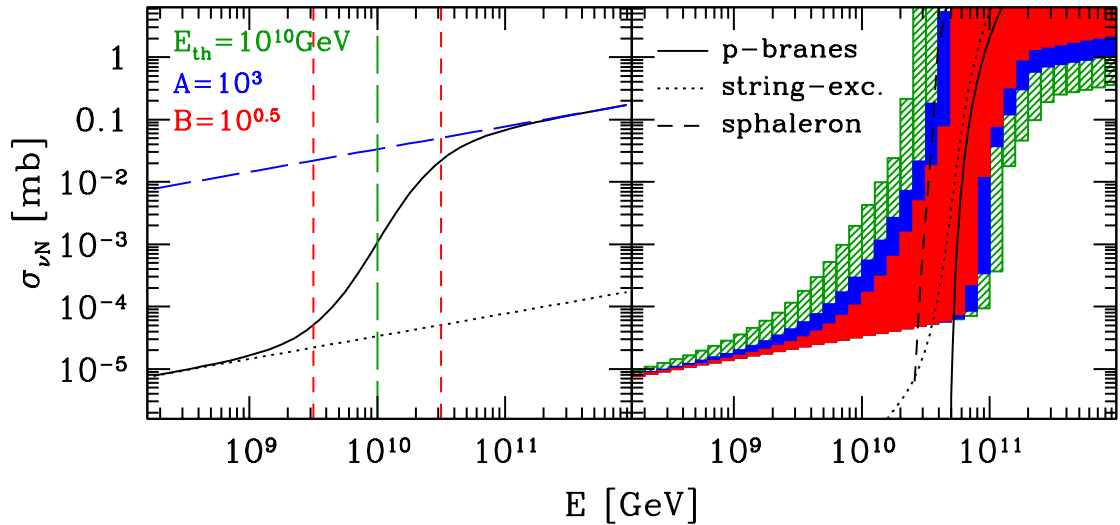} 
\caption{The range of the cross section within the 99\%, 95\% and 90\% CL. Also shown are theoretical predictions of the neutrino-nucleon cross-section enhanced by electroweak sphalerons, $p$-branes, and string excitations (see Ref.~\cite{Ahlers:2005zy}).}
\label{fig}
\end{minipage}
\end{figure}

\section{Strongly Interacting Neutrinos from Optically Thin Sources}

The flux of UHE extragalactic protons from distant sources is redshifted and also subject to $e^+e^-$ pair production and photopion-production in the CMB which can be taken into account by means of propagation functions. The resonantly produced photopions provide a {\it guaranteed} source of cosmogenic UHE neutrinos observed at Earth. In astrophysical accelerators inelastic scattering of the beam protons off the ambient photon gas in the source will also produce photopions which provide an additional source of UHE neutrinos. The corresponding spectrum will in general depend on the details of the source such as the densities of the target photons and the ambient gas~\cite{Mannheim:1998wp}. We have used the flux of CRs from {\it optically thin} sources using the luminosities given in Ref.~\cite{Ahlers:2005sn} in the goodness-of-fit test.

For a reasonable and consistent contribution of extragalactic neutrinos in vertical CRs one has to assume a strong and rapid enhancement of the neutrino nucleon interaction. The realization of such a behavior has been proposed in scenarios beyond the (perturbative) SM (see Ref.~\cite{Ahlers:2005zy}). For convenience, we have approximated the strong neutrino nucleon cross section in our analysis by a $\tanh$-behavior shown in Fig.~\ref{fig}, parameterized by the energy scale and width of the transition, and the amplification compared to the Standard Model predictions.

Our analysis showed that UHE CRs measured at AGASA and HiRes can be interpreted to the 90\% CL as a composition of extragalactic protons and strongly interacting neutrinos from optically thin sources in agreement with experimental results from horizontal events at AGASA and contained events at RICE (see Fig.~\ref{fig}). The Pierre Auger Observatory combines the experimental techniques of AGASA and HiRes as a hybrid detector. With a better energy resolution, much higher statistics and also stronger bounds on horizontal showers it will certainly help to clarify our picture of UHE CRs in the future.

\section*{Acknowledgements}
The author would like to thank the organizers of the {\sc Erice School on Nuclear
  Physics} 2005 {\it``Neutrinos in Cosmology, in Astro, Particle and Nuclear
  Physic''} for the inspiring workshop and VIHKOS ({\it``Virtuelles Institut f\"ur
  Hochenergiestrahlungen aus dem Kosmos''}) for support.

\end{document}